\documentclass[twocolumn,showpacs,amsmath,amssymb,prl]{revtex4}

\usepackage{graphicx}
\usepackage{dcolumn}
\usepackage{bm}
\usepackage{color}
\usepackage{ulem}
\usepackage{amsfonts,amsthm}

\begin{document}

\newcommand{\bR}{\mbox{\boldmath $R$}}
\newcommand{\tr}[1]{\textcolor{red}{#1}}
\newcommand{\trs}[1]{\textcolor{red}{\sout{#1}}}
\newcommand{\tb}[1]{\textcolor{blue}{#1}}
\newcommand{\tbs}[1]{\textcolor{blue}{\sout{#1}}}
\newcommand{\Ha}{\mathcal{H}}
\newcommand{\mh}{\mathsf{h}}
\newcommand{\mA}{\mathsf{A}}
\newcommand{\mB}{\mathsf{B}}
\newcommand{\mC}{\mathsf{C}}
\newcommand{\mS}{\mathsf{S}}
\newcommand{\mU}{\mathsf{U}}
\newcommand{\mX}{\mathsf{X}}
\newcommand{\sP}{\mathcal{P}}
\newcommand{\sL}{\mathcal{L}}
\newcommand{\sO}{\mathcal{O}}
\newcommand{\la}{\langle}
\newcommand{\ra}{\rangle}
\newcommand{\ga}{\alpha}
\newcommand{\gb}{\beta}
\newcommand{\gc}{\gamma}
\newcommand{\gs}{\sigma}
\newcommand{\vk}{{\bm{k}}}
\newcommand{\vq}{{\bm{q}}}
\newcommand{\vR}{{\bm{R}}}
\newcommand{\vQ}{{\bm{Q}}}
\newcommand{\vga}{{\bm{\alpha}}}
\newcommand{\vgc}{{\bm{\gamma}}}
\arraycolsep=0.0em
\newcommand{\Ns}{N_{\text{s}}}

\preprint{APS/123-QED}

\title{
Magnetic Properties of {\it Ab initio} Model for Iron-Based Superconductors LaFeAsO
}

\author{Takahiro Misawa}
\email{misawa@solis.t.u-tokyo.ac.jp}
\author{Kazuma Nakamura}%
\author{Masatoshi Imada}%
\affiliation{%
Department of Applied Physics, University of Tokyo, and JST CREST,
7-3-1 Hongo, Bunkyo-ku, Tokyo, 113-8656, Japan
}%

\date{\today}

\begin{abstract}
By using variational Monte Carlo method, we 
examine an effective low-energy model for LaFeAsO 
derived from an {\it ab initio} downfolding scheme. 
We show that quantum and many-body fluctuations 
{near a quantum critical point}
largely reduce 
the antiferromagnetic (AF) ordered moment and {the model not only 
quantitatively reproduces the 
small ordered moment 
in LaFeAsO, but also explains the diverse  
dependence on LaFePO, BaFe$_2$As$_2$ and FeTe.} 
We also find that LaFeAsO is 
under
large orbital fluctuations, 
sandwiched by the AF 
Mott insulator and weakly correlated metals.
The orbital fluctuations and Dirac-cone dispersion hold keys for
the diverse magnetic properties. 
\end{abstract}

\pacs{71.15.-m, 71.27.+a, 71.30.+h}
\maketitle

Discovery of iron-based superconductors has renewed interest on mechanisms of high-$T_c$ superconductivity~\cite{Kamihara_LaFeAsO}. 
Roles of electron correlations in the iron-based families and in the pairing mechanisms are under strong 
debates~\cite{Kuroki,Haule,QSi,Craco,Anisimov,Aichhorn,Hansmann,Liebsch}.  
An aspect common in these families is the nearby antiferromagnetic (AF) phases similarly to 
the cuprate high-$T_c$ superconductors. 
In the iron-based families, however, 
the AF orders are found in metals and show a variety of ordered moment 
ranging from $\sim 2$ $\mu_{{\rm B}}$ with a bicollinear order for FeTe to
0.3-0.6 $\mu_{{\rm B}}$ with an AF stripe (AFS) order for LaFeAsO~\cite{Mook_Nature2008,Qureshi}. 
Since 
{\it ab initio} 
density-functional calculations 
usually underestimate AF moments, 
it is unconventional that the measured moment for LaFeAsO is substantially smaller than 
the density-functional estimate of 1.2-2.6 $\mu_{{\rm B}}$~\cite{CaO_PRB2008,IshibashiTerakura,Ma_PRB2008,Mazin_PRB2008}. 
It strongly suggests the necessity of considering quantum fluctuations beyond the mean-field level.

A difficulty in the iron-based superconductors comes from an entangled band structure composed of 
five Fe-3$d$ bands near the Fermi level~\cite{Mazin_PRB2008}.  
Recently proposed three-stage scheme, consisting of the global band structure by 
the conventional density functional calculations, {\it ab initio} downfolding scheme to derive low-energy models, 
and solving the resultant realistic models,
has opened a way of analyzing such a real complexity of 
materials~\cite{Aryasetiawan_PRB2004,Imai_PRL2005}. 
This scheme has 
already been applied to the iron-based families and
effective low-energy models have been derived~\cite{NakamuraFeAs,Miyake}.
The next step of solving the effective models by reliable low-energy solvers is so far
mostly confined to that by the dynamical mean-field approximation (DMFA)~\cite{Haule,Aichhorn,Hansmann}, 
where spatial correlation effects are
hardly analyzed in the present multi-orbital systems.

To understand the correlation effects and the unconventional magnetism described above, 
interplays of orbitals and spins have to be elucidated by considering spatial fluctuations 
beyond DMFA.  
For this purpose, many-variable variational Monte Carlo (VMC) method~\cite{TaharaVMC_Letter}, 
combined with quantum-number projection~\cite{Mizusaki_PRB2004}, offers a suitable and accurate low-energy solver 
in clarifying dynamically and spatially 
fluctuating phenomena~\cite{TaharaVMC_Letter}. 

In this letter, we apply unrestricted 
Hartree-Fock (UHF) and VMC methods 
to solve
the downfolded effective low-energy model of LaFeAsO~\cite{NakamuraFeAs}.
We clarify the microscopic origin of the small AF ordered moment.
In addition, we find that LaFeAsO is located in a region of large orbital fluctuations, 
sandwiched by the AF Mott insulators and weakly correlated metals.

Our low-energy model 
derived from the {\it ab initio} scheme~\cite{NakamuraFeAs} is defined for ten-fold degenerate Fe-3$d$ 
orbitals in a unit cell containing two Fe atoms in the form
\begin{eqnarray}
\mathcal{H}
&=& \sum_{\sigma} \sum_{{\bm R} {\bm R'}} \sum_{nm}  
  t_{m {\bf R} n {\bf R}'} 
                   a_{n {\bm R}}^{\sigma \dagger} 
                   a_{m {\bm R'}}^{\sigma}   \nonumber \\
&+& \frac{1}{2} \sum_{\sigma \rho} \sum_{{\bm R}} \sum_{nm} 
  \biggl\{ U_{m {\bf R} n {\bf R}} 
                   a_{n {\bm R}}^{\sigma \dagger} 
                   a_{m {\bm R}}^{\rho \dagger}
                   a_{m {\bm R}}^{\rho} 
                   a_{n {\bm R}}^{\sigma} \nonumber \\ 
&+& J_{m {\bf R} n {\bf R}} 
\bigl(\!a_{n {\bm R}}^{\sigma \dagger} 
      \!a_{m {\bm R}}^{\rho \dagger}
      \!a_{n {\bm R}}^{\rho} 
      \!a_{m {\bm R}}^{\sigma} 
   \!+\!a_{n {\bm R}}^{\sigma \dagger} 
      \!a_{n {\bm R}}^{\rho \dagger}
      \!a_{m {\bm R}}^{\rho} 
      \!a_{m {\bm R}}^{\sigma}\bigr)\! \biggr\},
\label{Eq:Ham}
\end{eqnarray}
where $a_{n {\bm R}}^{\sigma \dagger}$ ($a_{n {\bm R}}^{\sigma}$) 
is a creation (annihilation) operator of an electron with 
spin $\sigma$ in the $n$th maximally localized Wannier orbitals~\cite{Marzari} 
centered on Fe atoms in the unitcell at $\bR$. 
$t_{m {\bf R} n {\bf R}'}$ contains single-particle levels and transfer integrals.
$U_{m {\bf R} n {\bf R}'}$ and 
$J_{m {\bf R} n {\bf R}'}$ are screened Coulomb and exchange 
interactions, respectively.  
Offsite interactions were dropped since those are more than four times smaller than the onsite parameters.
Details of the parameter derivation are given in Ref.~\cite{NakamuraFeAs}.

\begin{figure}[h!]
	\begin{center}
		\includegraphics[width=8.2cm,clip]{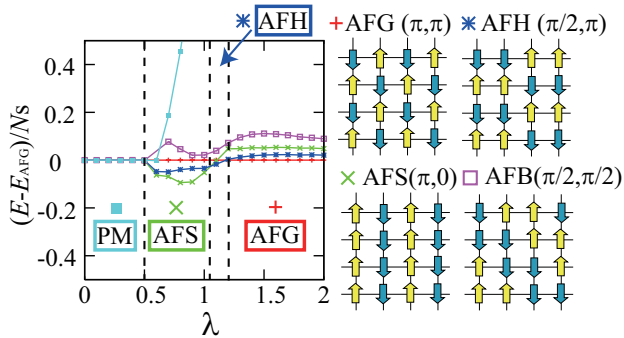}   
	\end{center}
\caption{(color online)~Left panel: Ground-state UHF energy per site of 
AF and paramagnetic metal (PM) as functions of 
interaction ratio $\lambda$.
Right panel: Four 
AF patterns. 
Energy is measured from that of AFG.
The ground state is a PM for $0<\lambda<0.5$, 
AFS for $0.5<\lambda<1.1$, and AFG for $\lambda>1.2$.
Around $\lambda\sim 1.1$, AFH is the ground state.
For $\lambda<0.5$, all the solutions converge to PM.}
\label{fig:Fig1}
\end{figure}%

Here, we introduce
a measure 
of the geometrical frustration
defined as a ratio of
the {diagonal} next-nearest transfer ($t^{\prime}$) to 
the nearest-neighbor transfer ($t$), i.e., $t^{\prime}/t$.
For $yz/zx$ and $x^{2}-y^{2}$ orbitals, 
this value is roughly
1.0, while for $xy$ and $z^{2}$ orbitals $t^{\prime}/t\sim 0.1$~\cite{transfer}.
Thus,
$yz/zx$ and $x^{2}-y^{2}$ orbitals are 
categorized to strongly frustrated orbitals whereas $xy$ and $z^{2}$ orbitals are unfrustrated.
This distinction in the degree of frustration is a characteristic feature of this model and controls
magnetic properties.
Another point to be stressed is an appreciable orbital dependence of the Coulomb interaction; the
 smallest intra-orbital Coulomb interaction is 
$U_{x^{2}-y^{2}}$ = 2.20 eV and the largest one is $U_{xy}$ = 3.31 eV. 
Such differences bring about different roles for different orbitals, 
namely an orbital differentiation by lifting the degeneracy,
in determining magnetic and charge structures. 

In the multi-band model,
effective one-body potential $\tilde{\mu}$ is given by
$\tilde{\mu}_{\nu}=\mu_{\nu}+\sum_{\nu^{\prime}}U_{\nu\nu^{\prime}}n_{\nu^{\prime}}+U_{\nu\nu}n_{\nu}/2$,
where 
$\mu_{\nu}$ is a one-body potential for orbital $\nu$ and 
$n_{\nu}$ is its occupation.
The Hartree term described as the second and third terms 
in $\tilde{\mu}_{\nu}$
is already included in the LDA calculations.
Therefore, 
we need to eliminate this Hartree contribution 
 to exclude the double counting.
To this end,
we correct $\mu_{\nu}$ so as to satisfy 
$\tilde{\mu}_{\nu}=\mu_{\nu}^{\rm LDA}$, where $\mu_{\nu}^{\rm LDA}$
is the one-body potential obtained from LDA.
For multi-band systems 
this correction is necessary. 
On the other hand, we ignore the double counting from the exchange correlation energy by LDA, 
because it is small~\cite{Anisimov97}. 

In order to monitor the Coulomb interactions $U$ as well as the
exchange interactions $J$,
we introduce the interaction ratio $\lambda$
to scale all the matrix elements of $U$ and $J$ uniformly. 
The original {\it ab initio} values (non-interacting case) correspond to $\lambda=1$ ($\lambda=0$).

To gain a rough insight into the effective Hamiltonian in Eq.~(\ref{Eq:Ham}),
we show results of UHF calculations in Fig.~\ref{fig:Fig1}.
Although the UHF overestimates stability of the ordered state, 
it is useful to understand the global structure
of the model.
We consider 
four different AF solutions
as shown in the right panel of Fig.~\ref{fig:Fig1}; 
 AFS, G-type AF state (AFG), and bicollinear AF state (AFB)
as the candidates for the ground states of the iron pnictides.
In addition,
we examine the stability of 
the half-collinear AF state (AFH) known to be realized in a frustrated Hubbard model~\cite{Mizusaki_PRB2004}.

In the left panel of Fig.~\ref{fig:Fig1}, we plot their energies as functions of $\lambda$. 
Here, $\mu_{\nu}$ is 
fixed at the value 
at
the original interaction parameter (
$\lambda=1$). 
With increasing $\lambda$, paramagnetic metal undergoes an AF phase transition around $\lambda=0.5$ into the AFS state.
The AFS phase continues to be stable through 
$\lambda=1$.
For $\lambda>1.2$, AFG becomes the ground state. 
Sandwiched by AFS and AFG, 
AFH is stabilized around $\lambda\sim1.1$, 
implying a relevant frustration effect in this region. 
The AFG phase is always the ground state 
when we ignore the double counting correction of the Hartree terms.

\begin{figure}[h!]
	\begin{center}
		\includegraphics[width=8.2cm,clip]{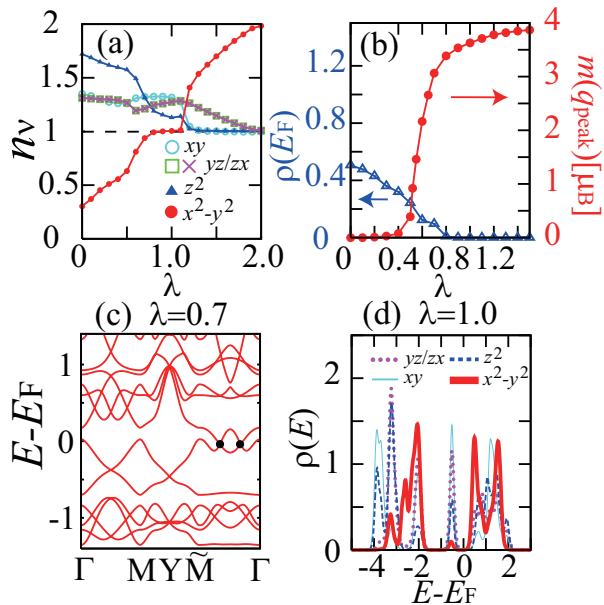}   
	\end{center}
\caption{(color online)~UHF results for our {\it ab initio} 
low-energy model: (a)~Orbital occupations as  functions of $\lambda$. 
Near $\lambda=1$, occupation of the $x^{2}-y^{2}$ orbital is pinned at 1. 
(b)~Magnetic ordered moment $m(q_{\rm peak})$ and
DOS at the Fermi level $\rho(E_{F})$. 
(c)~Band structures for $\lambda=0.7$ along the high-symmetry points, where $\Gamma=(0,0)$, ${\rm M}=(\pi,\pi)$,
${\rm Y}=(0,\pi)$, and $\tilde{{\rm M}}=(-\pi,\pi)$.
Energy zero is the Fermi level. 
Dirac nodes (dots)
appear along the $\tilde{{\rm M}}$-$\Gamma$ line.
(d)~Partial 
DOS at $\lambda=1.0$.
}
\label{fig:Fig2}
\end{figure}

Orbital occupations $n_{\nu}$ are 
monitored as functions of $\lambda$ 
in Fig.~\ref{fig:Fig2}~(a).
From this plot, we can understand 
why 
the AFS (AFG) phase is stable for $\lambda<1.2$ ($\lambda>1.2$):
For $\lambda\sim1$,
all the orbitals contribute to the magnetic moment;
each orbital occupation is
close to half filling ($n_{\nu}\sim1$). 
In this region, the number of the frustrated orbitals~($yz/zx$, $x^{2}-y^{2}$) is 
larger than 
that of the unfrustrated orbitals~($xy$, $z^{2}$).  
Hence, the former frustrated orbitals  
dominate the magnetic structure
and AFS becomes stable. 
In contrast, for $\lambda>1.2$, 
$n_{x^{2}-y^{2}}$ increases rapidly away from
half filling and 
loses magnetic activity.
Such a disruption in the subtle balance of the frustrated and unfrustrated orbitals leads the latter orbitals, having the stronger Coulomb 
interaction ($U_{xy}$ = 3.31 eV, $U_{z^{2}}$ = 3.27 eV) than the former ($U_{yz}$ = $U_{zx}$ = 2.77 eV), to dominate the magnetic structure;   
AFG is realized for large $\lambda$.
The result observed here is interpreted as {\it electron differentiation in orbitals}; 
the appreciable orbital dependence of the frustration and interaction parameters generates 
emergent distinction in roles of each orbital.

Figure \ref{fig:Fig2}~(b) plots $\lambda$ dependence of the magnetic ordered moment defined by
$m(\bm{q})^{2}={\frac{4}{3(N_{\rm s})^2} \sum_{i,j}\langle\bm{S}_{i}\bm{S}_{j}\rangle e^{i\bm{q(r_{i}-r_{j})}}}$
for $N_{\rm s}$-site ($N_{\rm s}/2$ unit cell) system with the periodic boundary condition and $\bm{q}$ is set to $q_{\rm peak}\equiv(0,\pi)$.
In this definition, the saturated magnetic moment for the classical N\'{e}el state is
given by $m$ = 4 $\mu_{\rm B}$.
We also superpose density of states (DOS) at the Fermi level [$\rho(E_{\rm F})$] in the same plot.
The AF phase transition occurs around $\lambda$ = 0.5 and 
then the metal-insulator transition occurs around $\lambda=0.8$.
The AF metal thus exists for $0.5\leq\lambda\leq0.8$.
Notice that the metal 
survives even for an unexpectedly large AF
moment,
$m\sim$ 3.5 $\mu_{\rm B}$
close to the saturated value.

Although real LaFeAsO corresponds to $\lambda\sim 0.5$ on the UHF level, 
it is illuminating to show UHF band structures at a large $\lambda=0.7$ 
near the metal-insulator transition to understand the origin of the robust metal
(see Fig.~\ref{fig:Fig2}~(c)). 
Near the Fermi level, 
two Dirac nodes (marked as dots) appear along 
the $\tilde{\rm M}$-$\Gamma$ line~\cite{Ran_PRB2008,DiracBaFe2As2}.
Until their pair annihilation, the Dirac nodes 
maintain the metallic band and the metal becomes stable
even when a large AF ordered moment
grows at a large $\lambda$ ($\sim 1$). 

In Fig.~\ref{fig:Fig2}~(d), we show partial 
DOS for $\lambda=1.0$ 
corresponding to the insulating region.
Just below the Fermi level,
the $x^{2}-y^{2}$ DOS  
is much smaller than those of the other orbitals, 
indicating a relatively large gap opening 
($\sim$ 2 eV) for this orbital.
This is consistent with the fact that, around $\lambda\sim1$, 
$n_{x^{2}-y^{2}}$ is pinned to half filling while the other orbital occupations change continuously [see Fig.~\ref{fig:Fig2}~(a)].
It has similarity to the orbital
selective Mott transitions~\cite{Koga_PRL2004}
in the sense that the charge gap 
depends largely on the orbitals,
although here, all the orbitals have nonzero charge gap. 

To examine how quantum fluctuations beyond the mean-field
approximation affect the magnetic properties, we 
performed 
VMC calculations.
Our variational wave function~\cite{TaharaVMC_Letter} is defined as
\begin{equation}
|\psi\ra = \sP_{\rm G}\sL^{S=0}\sL^{K=0}|\phi_{\rm pair}\ra,
\label{Eq:WF}
\end{equation}
where $\sP_{\rm G}$ is the Gutzwiller factor.
The spin (momentum) quantum-number projection $\sL^{S=0}$ ($\sL^{K=0}$) restores the
SU(2) spin-rotational (translational) symmetry and generates a state with the
correct total spin $S$=$0$ (total momentum $K$=$0$).
The one-body part $|\phi_{\rm pair}\ra$ is
the generalized pairing wave function defined as
$|\phi_{\rm pair}\ra=\Big[\sum_{i,j=1}^{\Ns}f_{ij}c_{i\uparrow}^{\dag}c_{j\downarrow}^{\dag}\Big]^{N/2} |0 \ra$
with $f_{ij}$ being the variational parameters.
In this study, we 
allow $f_{ij}$ to have $2\times2$ sublattice structure
or equivalently we have $2\times2\times5^{2}\times N_{\rm s}$ parameters.
All the variational parameters are simultaneously 
optimized by using the stochastic reconfiguration method~\cite{TaharaVMC_Letter,Sorella_PRB2001}.
The variational function $|\psi\ra$ 
in Eq.~(\ref{Eq:WF}) 
can describe paramagnetic metals and the AFG and AFS phases, as well as
superconducting phases.

\begin{figure}[h!]
	\begin{center}
		\includegraphics[width=6.5cm,clip]{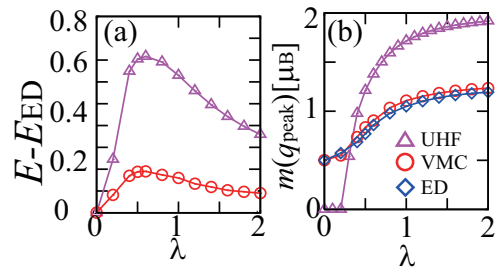}   
	\end{center}
\caption{(color online)~Benchmark of VMC on two-orbital 
model for $N_{\rm s}=4\times 2$.
(a)~$\lambda$ dependence of total energy obtained by UHF and VMC, measured from the exact result $E_{\rm ED}$. %
(b) 
$\lambda$ dependence of magnetic ordered moment.}
\label{fig:Fig3}
\end{figure}%

Before applying the VMC method to 
our model,
we show a benchmark demonstrating 
high accuracies of our VMC.
In LaFeAsO, because doubly degenerate $yz/zx$ 
orbitals have the largest weight on the Fermi surface,
we consider 
a two-orbital model for these orbitals 
with the same interactions
as the original five-orbital models.
This two-orbital model is strongly correlated and 
highly frustrated; i.e., $t^{\prime}/t\sim 1.0$ and $U/t\sim10$.

We compare in Fig.~\ref{fig:Fig3}~(a) UHF, VMC, and exact-diagonalization (ED) energies of an $N_{\rm s}$=4$\times$2 system.
The ground-state energy of VMC becomes much lower than 
the UHF energy and very close to the ED one.
Figure~\ref{fig:Fig3}~(b) illustrates that 
quantum and many-body fluctuations drastically decrease 
the ordered moment $m(q_{\rm peak})$ and 
VMC reproduces satisfactorily 
the ED result from the weak to strong coupling 
regimes.
Clearly, our VMC 
works well even
for serious regions of frustrations 
and electron correlations in the multi-orbital system.

\begin{figure}[h!]
	\begin{center}
		\includegraphics[width=8cm,clip]{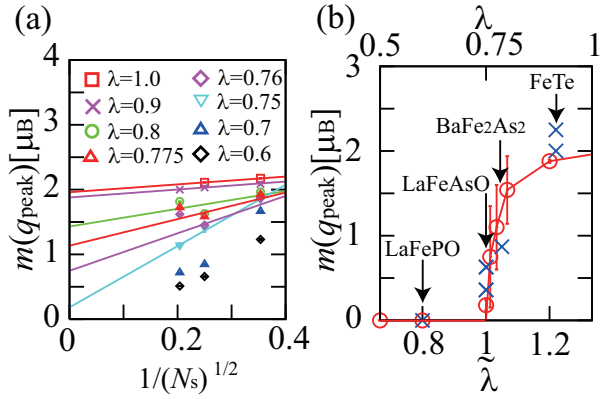}   
	\end{center}
\caption{(color online)~Magnetic ordered moment $m(q_{\rm peak})$ calculated by VMC for the model of LaFeAsO. 
(a) Size dependence of $m(q_{\rm peak})$ for several $\lambda$. 
(b) $\tilde{\lambda}$ dependence of $m(q_{\rm peak})$ in the thermodynamic limit (open circles). 
Experimentally observed materials dependence at corresponding $\tilde{\lambda}$ is also shown
by crosses.
 Quantum critical point of the AF transition appears at slightly below $\lambda=0.75 (\tilde{\lambda}\sim 1)$.
}
\label{fig:Fig4}
\end{figure}

We give in Fig.~\ref{fig:Fig4} the VMC results for the 
low-energy effective Hamiltonian for LaFeAsO.
The VMC ground state is found to be AFS for $\lambda\leq1$.
System size dependence of $m(q_{\rm peak})$ 
for various choices of $\lambda$ is 
shown in Fig.~\ref{fig:Fig4}~(a).  
We extrapolated the data by $1/\sqrt{N_{\rm s}}$,  
in which finite size effects are assumed to arise from linear spin-wave type
excitations in the AF ordered states.
The extrapolation gives $m>0$
only for $\lambda_{\rm c} > 0.75$,
while, for $\lambda < 0.75$, 
the ground state is paramagnetic.

Here, there exist two origins of the overestimate
of the interaction parameters derived in Ref.\cite{NakamuraFeAs}.
First, Ref.\cite{NakamuraFeAs} ignored the 
screening effects from La-$f$ electrons. 
This screening was shown to lead to the 
reduction of the average of orbital diagonal Coulomb 
interactions from $U_{\rm ave}$ to 0.9$U_{\rm ave}$~\cite{Miyake}.
The second is the interlayer screening effect, which has to be appended
when we employ purely two-dimensional models as in the present study. 
The screening by electrons on neighboring metallic layers    
reduces $U_{\rm ave}$ to 0.85$U_{\rm ave}$
for LaFeAsO~\cite{Nakamura10}. 
On closer inspection, 
$U_{\rm ave}$ decreases to $\sim 0.75U_{\rm ave}$ in total.
Therefore, to compare with the experimental results, 
we should simulate correlation effects by $ \tilde{\lambda}\equiv (U_{\rm ave}/\tilde{U}_{\rm ave})\lambda $ 
instead of $\lambda$, 
where $\tilde{U}_{\rm ave}$ is the corrected 
interaction 
$ \tilde{U}_{\rm ave}\sim 0.75U_{\rm ave} $.

The $\tilde{\lambda}$ dependence of $m(q_{\rm peak})$ 
is shown in Fig.~\ref{fig:Fig4}~(b).
It indicates that LaFeAsO is
close to the AF quantum critical point  
consistently with experiments~\cite{Wang_EPL2009}.
Although details of the model parameters are not fully implemented, 
the {\it ab initio} parameters~\cite{Miyake} tell us that other compounds are also simulated
roughly by the corresponding scaled ratio of $\tilde{\lambda}$ specified from the ratio of $\tilde{U}_{\rm ave}/\bar{t}$ with the averaged transfer $\bar{t}$ in Ref.\cite{Miyake}.
Estimated ordered moments for LaFePO, BaFs$_{2}$As$_{2}$
and FeTe are also all consistent with the 
experiments~\cite{Mook_Nature2008,Qureshi,Wang_EPL2009,Huang_PRL2008,Li_PRB2009,Bao_PRL2009}.
The relevance of quantum criticality is a recent subject of 
debates~\cite{Dai2009,CruzPRL2010,Luetkens2009}.
The present work supports its relevance.

Finally, we describe why quantum fluctuations largely reduce 
the ordered moment. 
Around $\lambda=1$,
each orbital occupation is at incommensurate filling,  
slightly away from half filling.
This 
is analogous to 
the doped Mott insulators 
when we consider each orbital occupation,
although the total filling is always commensurate.
As a result of this incommensurate filling, 
quantum fluctuations in orbitals
substantially destroy the AF ordered moment.


In summary, we have studied the magnetic properties of
the {\it ab initio} low-energy model for LaFeAsO.
State-of-the-art VMC calculations show  
that the realistic parameter stabilizes a stripe-type AF metal
close to a quantum critical point in agreement with the small ordered 
moment observed in the experiment.
Furthermore, a sharp $\tilde{\lambda}$ dependence of the ordered 
moment quantitatively explains the 
ordered moment of LaFePO, BaFe$_2$As$_2$ and FeTe as well on unified grounds.
Large fluctuations and differentiations in orbitals coexisting with
a robust metallic dispersion arising from the Dirac cones also hold the key 
to understanding the low-energy physics. 

\acknowledgements{The authors gratefully thank Daisuke Tahara for the use of his VMC
code and useful comments.}


\begin{thebibliography}{27}
\expandafter\ifx\csname natexlab\endcsname\relax\def\natexlab#1{#1}\fi
\expandafter\ifx\csname bibnamefont\endcsname\relax
  \def\bibnamefont#1{#1}\fi
\expandafter\ifx\csname bibfnamefont\endcsname\relax
  \def\bibfnamefont#1{#1}\fi
\expandafter\ifx\csname citenamefont\endcsname\relax
  \def\citenamefont#1{#1}\fi
\expandafter\ifx\csname url\endcsname\relax
  \def\url#1{\texttt{#1}}\fi
\expandafter\ifx\csname urlprefix\endcsname\relax\def\urlprefix{URL }\fi
\providecommand{\bibinfo}[2]{#2}
\providecommand{\eprint}[2][]{\url{#2}}

\bibitem[{\citenamefont{Kamihara et~al.}(2008)\citenamefont{Kamihara, Watanabe,
  Hirano, and Hosono}}]{Kamihara_LaFeAsO}
\bibinfo{author}{\bibfnamefont{Y.}~\bibnamefont{Kamihara} {\it et al.}},
  \bibinfo{journal}{J.\ Am.\ Chem.\ Soc} \textbf{\bibinfo{volume}{130}},
  \bibinfo{pages}{3296} (\bibinfo{year}{2008}).


\bibitem[{\citenamefont{Kuroki et~al.}(2008)\citenamefont{Kuroki, Onari, Arita,
  Usui, Tanaka, Kontani, and Aoki}}]{Kuroki}
\bibinfo{author}{\bibfnamefont{K.}~\bibnamefont{Kuroki} {\it et al.}},
  \bibinfo{journal}{Phys. Rev. Lett.} \textbf{\bibinfo{volume}{101}},
  \bibinfo{pages}{087004} (\bibinfo{year}{2008}).

\bibitem[{\citenamefont{Haule et~al.}(2008)\citenamefont{Haule, Shim, and
  Kotliar}}]{Haule}
  \bibinfo{author}{\bibfnamefont{K.}~\bibnamefont{Haule} {\it et al.}},
  \bibinfo{journal}{Phys. Rev. Lett.} \textbf{\bibinfo{volume}{100}},
  \bibinfo{pages}{226402} (\bibinfo{year}{2008}).

{
\bibitem{QSi}
Q. Si and E. Abrahams,
Phys. Rev. Lett. {\bf 101}, 076401 (2008). 
}

{
\bibitem{Craco}
L. Craco {\it et al.},
Phys. Rev. B {\bf 78}, 134511 (2008). 
}


\bibitem[{\citenamefont{Anisimov et~al.}(2009)\citenamefont{Anisimov, Kurmaeva,
  Moewesb, and Izyumov}}]{Anisimov}
\bibinfo{author}{\bibfnamefont{V.~I.} \bibnamefont{Anisimov} {\it et al.}},
  \bibinfo{journal}{Physica C} \textbf{\bibinfo{volume}{469}},
  \bibinfo{pages}{442} (\bibinfo{year}{2009}).

\bibitem[{\citenamefont{Aichhorn et~al.}(2009)\citenamefont{Aichhorn,
  Pourovskii, Vildosola, Ferrero, Parcollet, Miyake, Georges, and
  Biermann}}]{Aichhorn}
\bibinfo{author}{\bibfnamefont{M.}~\bibnamefont{Aichhorn} {\it et al.}},
  \bibinfo{journal}{Phys. Rev. B} \textbf{\bibinfo{volume}{80}},
  \bibinfo{pages}{085101} (\bibinfo{year}{2009}).

\bibitem[{\citenamefont{Hansmann et~al.}()\citenamefont{Hansmann, Arita,
  Toschi, Sakai, Sangiovanni, and Held}}]{Hansmann}
\bibinfo{author}{\bibfnamefont{P.}~\bibnamefont{Hansmann} {\it et al.}},
  \bibinfo{howpublished}{arXiv:1003.2162v1}.

{
\bibitem{Liebsch}
H. Ishida and A. Liebsch,
Phys. Rev. B {\bf 81}, 054513 (2010). 
}

\bibitem[{\citenamefont{de~la Cruz et~al.}(2008)\citenamefont{de~la Cruz,
  Huang, Lynn, Li, II, Zarestky, Chen, Luo, Wang, and Dai}}]{Mook_Nature2008}
\bibinfo{author}{\bibfnamefont{C.}~\bibnamefont{de~la Cruz} {\it et al.}},
  \bibinfo{journal}{Nature} \textbf{\bibinfo{volume}{453}},
  \bibinfo{pages}{899} (\bibinfo{year}{2008}).

\bibitem[{\citenamefont{Qureshi et~al.}(2009)\citenamefont{N. Qureshi, 
Y. Drees, J. Werner, S. Wurmehl, C. Hess, R. Klingeler, B. Buechner, M. T. Fernandez-Diaz, M. Braden}}]{Qureshi}
  \bibinfo{author}{\bibfnamefont{N.}~\bibnamefont{Qureshi} {\it et al.}}, 
  \bibinfo{howpublished}{arXiv:1002.4326v1}.

\bibitem[{\citenamefont{Cao et~al.}(2008)\citenamefont{Cao, Hirschfeld, and
  Cheng}}]{CaO_PRB2008}
\bibinfo{author}{\bibfnamefont{C.}~\bibnamefont{Cao} {\it et al.}},
  \bibinfo{journal}{Phys. Rev. B} \textbf{\bibinfo{volume}{77}},
  \bibinfo{pages}{220506(R)} (\bibinfo{year}{2008}).

\bibitem[{\citenamefont{Ishibashi et~al.}(2008)\citenamefont{Ishibashi,
  Terakura, and Hosono}}]{IshibashiTerakura}
\bibinfo{author}{\bibfnamefont{S.}~\bibnamefont{Ishibashi} {\it et al.}},
  \bibinfo{journal}{J.\ Phys.\ Soc.\ Jpn} \textbf{\bibinfo{volume}{77}},
  \bibinfo{pages}{053709} (\bibinfo{year}{2008}).

\bibitem[{\citenamefont{Ma and Lu}(2008)}]{Ma_PRB2008}
\bibinfo{author}{\bibfnamefont{F.}~\bibnamefont{Ma}} \bibnamefont{and}
  \bibinfo{author}{\bibfnamefont{Z.-Y.} \bibnamefont{Lu}},
  \bibinfo{journal}{Phys. Rev. B} \textbf{\bibinfo{volume}{78}},
  \bibinfo{pages}{033111} (\bibinfo{year}{2008}).

\bibitem[{\citenamefont{Mazin et~al.}(2008)\citenamefont{Mazin, Johannes,
  Boeri, Koepernik, and Singh}}]{Mazin_PRB2008}
\bibinfo{author}{\bibfnamefont{I.~I.} \bibnamefont{Mazin} {\it et al.}},
  \bibinfo{journal}{Phys. Rev. B} \textbf{\bibinfo{volume}{78}},
  \bibinfo{pages}{085104} (\bibinfo{year}{2008}).

\bibitem[{\citenamefont{Aryasetiawan et al.}(2004)}]{Aryasetiawan_PRB2004}
\bibinfo{author}{\bibfnamefont{F.}~\bibnamefont{Aryasetiawan} {\it et al.}},
  \bibinfo{journal}{Phys.\ Rev.\ B} \textbf{\bibinfo{volume}{70}},
  \bibinfo{pages}{195104} (\bibinfo{year}{2004}).

\bibitem[{\citenamefont{Imai et~al.}(2005)\citenamefont{Imai, Solovyev, and
  Imada}}]{Imai_PRL2005}
\bibinfo{author}{\bibfnamefont{Y.}~\bibnamefont{Imai} {\it et al.}},
  \bibinfo{journal}{Phys.\ Rev.\ Lett.} \textbf{\bibinfo{volume}{95}},
  \bibinfo{pages}{176405} (\bibinfo{year}{2005});
  \bibinfo{author}{\bibfnamefont{Y.}~\bibnamefont{Imai} {\it et al.},} 
  \bibinfo{journal}{J.\ Phys.\ Soc.\ Jpn} \textbf{\bibinfo{volume}{75}},
  \bibinfo{pages}{094713} (\bibinfo{year}{2006});
  \bibinfo{author}{\bibfnamefont{Y.}~\bibnamefont{Otsuka} {\it et al.},} 
  \bibinfo{journal}{J.\ Phys.\ Soc.\ Jpn} \textbf{\bibinfo{volume}{75}},
  \bibinfo{pages}{124707} (\bibinfo{year}{2006}).



\bibitem[{\citenamefont{Nakamura et~al.}(2008)\citenamefont{Nakamura, Arita,
  and Imada}}]{NakamuraFeAs}
\bibinfo{author}{\bibfnamefont{K.}~\bibnamefont{Nakamura} {\it et al.}},
  \bibinfo{journal}{J.\ Phys.\ Soc.\ Jpn} \textbf{\bibinfo{volume}{76}},
  \bibinfo{pages}{4510} (\bibinfo{year}{2008}).

\bibitem[{\citenamefont{Miyake et~al.}()\citenamefont{Miyake, Nakamura, Arita,
  and Imada}}]{Miyake}
\bibinfo{author}{\bibfnamefont{T.}~\bibnamefont{Miyake} {\it et al.}},
  \bibinfo{journal}{J.\ Phys.\ Soc.\ Jpn.} \textbf{\bibinfo{volume}{79}},
  \bibinfo{pages}{044705} (\bibinfo{year}{2010}).

\bibitem[{\citenamefont{Tahara and
  Imada}(2008{\natexlab{a}})}]{TaharaVMC_Letter}
\bibinfo{author}{\bibfnamefont{D.}~\bibnamefont{Tahara}} \bibnamefont{and}
  \bibinfo{author}{\bibfnamefont{M.}~\bibnamefont{Imada}}, \bibinfo{journal}{J.
  Phys. Soc. Jpn} \textbf{\bibinfo{volume}{77}}, \bibinfo{pages}{093703}
  (\bibinfo{year}{2008}{\natexlab{a}});
\bibinfo{journal}{J. Phys. Soc. Jpn} \textbf{\bibinfo{volume}{77}}, \bibinfo{pages}{114701}
  (\bibinfo{year}{2008}{\natexlab{b}}).



\bibitem[{\citenamefont{Mizusaki and Imada}(2004)}]{Mizusaki_PRB2004}
\bibinfo{author}{\bibfnamefont{T.}~\bibnamefont{Mizusaki}} \bibnamefont{and}
  \bibinfo{author}{\bibfnamefont{M.}~\bibnamefont{Imada}},
  \bibinfo{journal}{Phys. Rev. B} \textbf{\bibinfo{volume}{69}},
  \bibinfo{pages}{125110} (\bibinfo{year}{2004});
  \bibinfo{journal}{Phys. Rev. B} \textbf{\bibinfo{volume}{74}},
  \bibinfo{pages}{014421} (\bibinfo{year}{2006}).


\bibitem[{\citenamefont{Marzari and Vanderbilt}(1997)}]{Marzari}
\bibinfo{author}{\bibfnamefont{N.}~\bibnamefont{Marzari}} \bibnamefont{and}
  \bibinfo{author}{\bibfnamefont{D.}~\bibnamefont{Vanderbilt}},
  \bibinfo{journal}{Phys. Rev. B} \textbf{\bibinfo{volume}{56}},
  \bibinfo{pages}{12847} (\bibinfo{year}{1997}); 
I. Souza {\it et al.}, Phys. Rev. B {\bf 65} 035109 (2001).

\bibitem[{tra()}]{transfer}
\bibinfo{note}{{Maximum diagonal next-nearest transfers are given by
  $|t_{xy,xy}^{\prime}|=0.06$,
  $|t_{yz,yz}^{\prime}|=|t_{zx,zx}^{\prime}|=0.33$,
  $|t_{z^{2},z^{2}}^{\prime}|=0.00$, and
  $|t_{x^{2}-y^{2},x^{2}-y^{2}}^{\prime}|=0.13$ eV. Notice that transfers of
  $yz$ and $zx$ orbitals denpend on orientation. Non-vanishing off-diagonal
  elements are $|t_{xy,yz}^{\prime}|=0.14$, $|t_{z^{2},zx}^{\prime}|=0.14$, and
  $|t_{z^{2},x^{2}-y^{2}}^{\prime}|=0.18$ eV. Values of further neighbor
  transfers are found in Ref.~\cite{Miyake}. In the present calculations, we
  employed transfers up to the fourth neighbors.}}

{\bibitem[{\citenamefont{Anisimov97}(1997)}]{Anisimov97}
An estimate of the double counting selfenergy from a standard theory by A.I. Anisimov {\it et al.}
J. Phys. Condens. Matter, 9, 767 (1997) gives
$\Sigma_{DC} = \Sigma_U-\Sigma_J$,
consisting of the Hartree term $\Sigma_U = U(n-1/2)$, and the exchange term $\Sigma_J = J(n/2-1/2)$ with the estimate by the Slater integral as 
$U=2.1$ eV and $J=0.65$ eV.  The density $n=6$ gives $\Sigma_U\sim 11.6$eV $\gg \Sigma_J\sim 1.63$ eV. 
In addition, the selfenergy to be subtracted is only the orbital dependent part, which is especially small for the exchange part 
because of the orbital insensitive exchange interaction. 
}

\bibitem[{\citenamefont{Ran et~al.}(2009)\citenamefont{Ran, Wang, Zhai,
  Vishwanath, and Lee}}]{Ran_PRB2008}
\bibinfo{author}{\bibfnamefont{Y.}~\bibnamefont{Ran} {\it et al.}},
  \bibinfo{journal}{Phys. Rev. B} \textbf{\bibinfo{volume}{79}},
  \bibinfo{pages}{014505} (\bibinfo{year}{2009}).

\bibitem{DiracBaFe2As2}
P. Richard {\it et al.},
Phys. Rev. Lett. {\bf 104}, 137001 (2010). 

\bibitem[{\citenamefont{Koga et~al.}(2004)\citenamefont{Koga, Kawakami, Rice,
  and Sigrist}}]{Koga_PRL2004}
\bibinfo{author}{\bibfnamefont{A.}~\bibnamefont{Koga} {\it et al.}},
  \bibinfo{journal}{Phys. Rev. Lett.} \textbf{\bibinfo{volume}{92}},
  \bibinfo{pages}{216402} (\bibinfo{year}{2004}).

\bibitem[{\citenamefont{Sorella}(2001)}]{Sorella_PRB2001}
\bibinfo{author}{\bibfnamefont{S.}~\bibnamefont{Sorella}},
  \bibinfo{journal}{Phys. Rev. B} \textbf{\bibinfo{volume}{64}},
  \bibinfo{pages}{024512} (\bibinfo{year}{2001}).

\bibitem[{\citenamefont{Nakamura10}(2010)}]{Nakamura10}
\bibinfo{author}{\bibfnamefont{K.}~\bibnamefont{Nakamura} {\it et al.}},
  \bibinfo{journal}{unpublished.} 

\bibitem[{\citenamefont{Wang et~al.}(2009)\citenamefont{Wang, Jiang, Tao, Ren,
  Li, Li, Feng, Dai, Cao, and Xu}}]{Wang_EPL2009}
  \bibinfo{author}{\bibfnamefont{C.}~\bibnamefont{Wang} {\it et al.}}, 
  \bibinfo{journal}{Euro.\ Phys.\ Lett.} \textbf{\bibinfo{volume}{86}},
  \bibinfo{pages}{47002} (\bibinfo{year}{2009}).

\bibitem{Huang_PRL2008}
Q. Huang {\it et al.},
Phys. Rev. Lett. {\bf 101}, 257003 (2008). 

\bibitem{Li_PRB2009}
S. Li {\it et al.},
Phys. Rev. B {\bf 79}, 054503 (2009). 

\bibitem{Bao_PRL2009}
W. Bao {\it et al.},
Phys. Rev. Lett. {\bf 102}, 247001 (2009).

\bibitem{Dai2009}
J. Dai {\it et al.},
Proc. Nat. Acad. Sci. {\bf 106}, 4118 (2009). 

\bibitem{CruzPRL2010}
C. de la Cruz {\it et al.},
Phys. Rev. Lett. {\bf 104}, 017204 (2010). 

\bibitem{Luetkens2009}
H. Luetkens {\it et al.},
Nat. Mat. {\bf 8}, 305 (2009). 

\end{thebibliography}

\end{document}